\documentclass[aps,groupedaddress,nofootinbib,showkeys,showpacs,amsfonts,eqsecnum]{revtex4}
\usepackage{bm}
\usepackage{amsmath}
\newcommand{\tr}{{\rm tr \,}}
\newcommand{\Tr}{{\rm Tr}}
\newcommand{\Det}{{\rm Det\,}}
\newcommand{\Dirac}{{\bf D}}
\newcommand{\Kop}{{\bf K}}

\newcommand{\thru}[1]{\mathrel{\mathop{#1\!\!\!\!/}}}
\newcommand{\thrur}[1]{\mathrel{\mathop{#1\!\!\!/}}}

\newcommand{\D}{D}
\newcommand{\R}{Q}

\newcommand{\mlr}{m_{LR}}
\newcommand{\mrl}{m_{RL}}

\begin{document} 

\title{The invariant factor of the chiral determinant}

\author{L. L. Salcedo}

\affiliation{ Departamento de F{\'\i}sica At\'omica, Molecular y
Nuclear, Universidad de Granada, E-18071 Granada, Spain }

\date{\today}

\begin{abstract}
  The coupling of spin 0 and spin 1 external fields to Dirac fermions
  defines a theory which displays gauge chiral symmetry. Quantum
  mechanically, functional integration of the fermions yields the
  determinant of the Dirac operator, known as the chiral determinant.
  Its modulus is chiral invariant but not so its phase, which carries
  the chiral anomaly through the Wess-Zumino-Witten term. Here we find
  the remarkable result that, upon removal from the chiral determinant
  of this known anomalous part, the remaining chiral invariant factor
  is just the square root of the determinant of a local covariant
  operator of the Klein-Gordon type. This procedure bypasses the
  integrability obstruction allowing to write down a functional that
  correctly reproduces both the modulus and the phase of the chiral
  determinant. The technique is illustrated by computing the effective
  action in two dimensions at leading order in the derivative
  expansion. The results previously obtained by indirect methods are
  indeed reproduced.
\end{abstract}



\maketitle

\section{Introduction}
\label{sec:1}

In this work we consider even dimensional Dirac fermions which move in
the presence of external fields of the type scalar, pseudo-scalar,
vector and axial-vector and with general non abelian internal degrees
of freedom. At the classical level the theory is invariant under
chiral gauge transformations.  Quantum-mechanically, Feynman's
functional integral has to be evaluated, formally producing the
determinant of the Dirac operator, the so called chiral determinant.
There is large amount of literature on this subject. A good
recollection of it can be found in \cite{Ball:1989xg}. Much work has
been devoted to the purely gauge case (a single chirality and no zero
spin external fields) regarding its consistency and anomalies. Here we
will discuss the full coupling case only. This kind of setting appears
naturally in low energy quark models of QCD
\cite{Gasser:1983yg,Espriu:1989ff,Christov:1995vm,Bijnens:1995ww} and
so has direct phenomenological implications. More generally it is of
immediate interest in models where fermions have to be integrated out.
See e.g. \cite{Smit:2004kh} for an application to the study of CP
violation in early cosmology.

The effective action functional, the logarithm of the chiral
determinant, is an extension of the thermodynamic potentials where the
Lagrangian multipliers are external background fields. This functional
compactly embodies the properties of the field-theoretic system and in
particular its symmetries \cite{Itzykson:1980bk}.

As is well-known chiral symmetry is not preserved after quantization,
rather it develops an anomaly which survives the process of regulating
the ultraviolet
divergencies\cite{Adler:1969gk,Bell:1969ts,Bardeen:1969md} . The
chiral anomaly does not affect the real part of the effective action
\cite{AlvarezGaume:1983ig}; this quantity can be mapped to a bosonic
theory and so can be computed using a plethora of techniques since
chiral invariance can be invoked to simplify the calculations.

The imaginary part of the effective action carries the chiral anomaly.
For this reason it is more difficult to work with but also more
interesting.  Most efforts have concentrated on the properties of the
anomaly and of its generating effective action, the gauged
Wess-Zumino-Witten term, since it poses a theoretical challenge
\cite{Adler:1969er,Wess:1971yu,Fujikawa:1979ay,Goldstone:1981kk,%
  Witten:1983tw,Kaymakcalan:1983qq,Bardeen:1984pm,D'Hoker:1984ph,%
  Leutwyler:1985em,Banerjee:1986bu,Ball:1986ic,Ball:1989xg,%
  Alvarez-Gaume:1985dr}.  The computation of the effective action
itself has received less attention. In some cases, such as scalar and
pseudo-scalar fields complying with the chiral circle constraint, the
imaginary part of the effective action is saturated by the
Wess-Zumino-Witten term, at least to leading order, but in the general
case, or at finite temperature \cite{Salcedo:1998tg}, there is a non
trivial chiral invariant remainder.

The calculation of the imaginary part of the effective action is
complicated by the lack of chiral invariance, both in the functional
itself and in the formalism used to compute it. Perturbative
calculations or inverse mass expansions hide the underlying chiral
invariance of the remainder once the Wess-Zumino-Witten term has been
separated.  In this view the derivative expansion method is
advantageous. It is a non-perturbative approach which has the virtue
that different orders are not mixed by chiral transformations. In the
imaginary part, the expansion starts with the term with as many
derivatives as the dimension of the space-time. This leading order
term is the only one affected by ultraviolet divergencies, and hence
the only one with anomalous breaking of the chiral symmetry.

However, even if the expansion itself preserves chiral symmetry, this
symmetry can be spoiled by the regularization. This is the case of
regularizations such as the heat kernel of the squared Dirac operator,
$\Dirac^2$ (a Klein-Gordon like operator) \cite{Dhar:1985gh} or direct
$\zeta$-function regularization of the $\Dirac$
\cite{GamboaSaravi:1983xw,Salcedo:1996qy} or $\Dirac^2$. This is
because chiral symmetry does not act as a similarity transformation on
the Dirac operator, and so powers of this operator do transform in a
complicated way.

In order to have a proper construction of the effective action in the
anomalous sector, it is natural to use the current, i.e., the
variation of the effective action under a gauge field deformation.
The current has a chiral covariant part which is therefore amenable to
simple and direct computation. There is an integrability obstruction,
however: as a consequence of the anomaly such current is not directly
consistent \cite{Leutwyler:1985em,Ball:1989xg}; it differs from the
covariant one by a known counter-term \cite{Bardeen:1984pm}.

Once the obstruction is bypassed and the consistent current is
obtained, the effective action in the abnormal parity sector gets well
defined \cite{Ball:1989xg}. This idea has been successfully
implemented in \cite{Salcedo:2000hx} where the method of covariant
symbols \cite{Pletnev:1998yu,Salcedo:2006pv} is used to obtain the
covariant current.  There the leading order term was obtained in two
and four dimensions.  The same approach has been applied (this time
using the world-line method instead of covariant symbols) in
\cite{Hernandez:2007ng} to compute the next-to-leading order in two
dimensions.

The motivation for this work is as follows: Upon removal from the
fermionic effective action of its well understood anomalous
contribution, a chiral invariant functional is obtained. However, this
observation does not directly provide us with a set of ``Feynman
rules'' to carry out manifestly chiral covariant calculations. The
procedure based on the current, to bypass the integrability
obstruction, is indirect.  On the other hand, all ultraviolet finite
contributions are free from any anomaly and so they would not be
affected by any related obstruction.  The latter should only affect
the Wess-Zumino-Witten term.  Therefore, if one wanted to compute the
chiral invariant remainder at leading order in the derivative
expansion, or the effective action beyond leading order, or finite
temperature corrections (thermal corrections are ultraviolet and the
anomaly is temperature independent \cite{GomezNicola:1994vq}), etc,
there should be no obstruction from the anomaly.  This suggests that
such chiral invariant calculational scheme should exists. The Feynman
rules-like scheme follows after the chiral invariant part of the
effective action can be codified in the form $\Tr\log\Kop$.  This is
precisely what is achieve here. Namely, we construct a second order
differential operator, $\Kop$ (cf. (\ref{eq:main})) that is manifestly
chiral covariant and such that $(\Det\Kop)^{1/2}$ is just the chiral
invariant part of the chiral determinant (modulus and phase). Since
$\Kop$ is a standard Klein-Gordon operator, essentially the same
methods available to the real part are applicable here. We illustrate
our result by computing the effective action to two derivatives in a
strict derivative expansion (no other approximation is involved) in
two dimensions, and show that the correct result is reproduced.

Section \ref{sec:2a} indtroduces definitions and summarizes results
related to the chiral determinant. Section \ref{sec:3a} describes the
construction of the Klein-Gordon operator accounting for the chiral
invariant part of the effective action and proves the main result. An
application is also presented within the derivative expansion. Section
\ref{sec:conclusions} presents the conclusions.

\section{The chiral determinant}
\label{sec:2a}

In this section we summarize some theoretical results in
the literature regarding chiral fermions and their effective action.

\subsection{The Dirac operator}

We consider a Dirac operator $\Dirac$ describing Dirac fermions
coupled to spin 0 and spin 1 external fields with non abelian degrees
of freedom:
\begin{equation}
\Dirac = \gamma_\mu(\partial_\mu+{\cal V}_\mu)
+\gamma_\mu\gamma_5 {\cal A}_\mu
+{\cal S}
+\gamma_5 {\cal P} .
\end{equation}
The fermions live in a $d$-dimensional Euclidean space-time, and $d$
is even. We will only consider in this work the case of zero
temperature and flat space-time. Our conventions are: 
\begin{equation}
\gamma_\mu= \gamma_\mu^\dagger\,,
\qquad
\{\gamma_\mu,\gamma_\nu\}=2\delta_{\mu\nu}
\,,
\qquad
\gamma_5 = \gamma_5^\dagger = i^{d/2}\gamma_0\cdots\gamma_{d-1}
\,,
\qquad
\tr_{\text{Dirac}}(1)=2^{d/2}
 \,.
\end{equation}
The external fields ${\cal V}_\mu(x)$, ${\cal A}_\mu(x)$, ${\cal
  S}(x)$ and ${\cal P}(x)$ are square matrices in the space of the fermionic
internal degrees of freedom. Unitarity requires ${\cal S}(x)$ to be a
hermitian matrix and ${\cal V}_\mu(x)$, ${\cal A}_\mu(x)$ and ${\cal
  P}(x)$ to be antihermitian.

In order to emphasize the chiral properties it will be convenient to
work with fields with well defined transformation under chiral
rotations. To this end we express the Dirac operator in the
form:\footnote{$R$ and $L$  stand for right
  and left chirality respectively. We put them as sub- or
  super-indices indistinguishably.}
\begin{equation}
\Dirac= \thru{\D}_R P_R +\thru{\D}_L P_L + \mlr P_R+\mrl P_L \,,
\label{eq:2.3}
\end{equation}
where 
\begin{equation}
P_R= \frac{1}{2}(1+\gamma_5) \,,\quad
P_L= \frac{1}{2}(1-\gamma_5) \,,\quad
\end{equation}
and
\begin{equation}
D^{R,L}_\mu=\partial_\mu+v^{R,L}_\mu
\,,\qquad
v^{R,L}_\mu =  {\cal V}_\mu \pm  {\cal A}_\mu
\,,\qquad
\mlr=  {\cal S}+ {\cal P}
\,,\qquad
\mrl =  {\cal S}- {\cal P}
\,.
\end{equation}
Unitarity then requires
\begin{equation}
(v^{R,L}_\mu)^\dagger = - v^{R,L}_\mu
\,,\qquad
\mlr^\dagger= \mrl
\,.
\end{equation}
In addition, we assume the matrices $\mlr$, $\mrl$ to be nowhere
singular. This excludes the much studied case of fermions with a
single chirality, but avoids infrared singularities (in particular,
in the derivative expansion to be considered below).

\subsection{The effective action}

The fermionic effective action $W$ is introduced through standard
functional integration of the fermionic fields
\begin{equation}
e^{-W}=\int {\cal D}\bar\psi {\cal D}\psi 
\,e^{-\int d^dx \,\bar\psi \Dirac \psi} 
= \Det \Dirac
\end{equation}
so formally
\begin{equation}
W= -\Tr\,\log\Dirac
\end{equation}
modulo ultraviolet (UV) ambiguities. $\Tr$ denotes the functional
trace and includes space-time, internal and Dirac degrees of freedom.

It is important to recall that the Dirac operator (or more generally an
action) does not define a single quantum field theory but a whole
class of them.  The UV ambiguities affecting the effective action can
be exposed, e.g., through its computation within perturbation theory.
$W$ represents the sum of all one-loop Feynman diagrams, where the
fermion runs over the loop and the external fields correspond to
insertions in the loop. All diagrams with more than $d$ such
insertions have more than $d$ fermion propagators and so are UV
finite. These contributions are therefore independent of how the
theory is regularized and renormalized. The UV ambiguity is thus a
polynomial of the external fields of degree at most
$d$.\footnote{Alternatively, taking $n$ successive variations of the
  effective action with respect the external fields in $\Dirac$, with
  $n>d$, yields an operator of the type $\Dirac^{-n}$ whose trace is
  UV convergent.}  Likewise, taking a derivative with respect the
momentum of the insertion decreases the UV divergence degree by one.
This implies that the UV ambiguity is a polynomial in the external
momenta. In summary, the UV ambiguity in $W$ is just a local
Lagrangian polynomial of the external fields and their derivatives, of
mass dimension $d$, i.e., the standard counter-term allowed in a
renormalizable theory. (Note that in principle one can choose to
include in the counter-term new external fields not present in the
original theory and this is advisable in some circumstances
\cite{Bijnens:1993cy,RuizArriola:1995ea}.) So the rule is to compute
$W$ using any valid procedure (i.e., one that preserves all UV finite
contributions) and then add the appropriate counter-term to obtain any
of the several theories described by the same Dirac operator. That is,
if $W_0(\Dirac)$ is a renormalized effective action, any other
determination of the effective action is related to this one by the
relation
\begin{equation}
W(\Dirac)= W_0(\Dirac)+W_{\text{ct}}(\Dirac) 
\label{eq:2a.9}
\end{equation}
where $W_{\text{ct}}=\int d^d x\,{\cal L}_{\text{ct}}(x)$ and ${\cal
  L}_{\text{ct}}(x)$ is a polynomial of degree at most $d$ in the
``variables'' $\partial_\mu$, $v_R$, $v_L$, $\mlr$, and $\mrl$.

One of the ways to achieve a valid definition of $W(\Dirac)$ is
through the $\zeta$-function technique \cite{Hawking:1977ja}. The
function
\begin{equation}
\zeta(s,\Dirac)= \Tr(\Dirac^s)
\end{equation}
is UV finite for ${\rm Re\,} s<-d$. Its analytical extension is a
meromorphic function in the complex plane $s$ with simple poles at
$s=-1,-2,\ldots,-d$ \cite{Seeley:1967ea}.  Because $s=0$ is a regular
point, the effective action can be defined as
\begin{equation}
W(\Dirac)=-\frac{d}{ds}\Tr(\Dirac^s) \Big|_{s=0} \,.
\end{equation}
Note that no further renormalization is needed, as the right-hand side
is already UV finite. An interesting property of this renormalization
is that it only depends on the spectrum of $\Dirac$: let $\lambda_n$
be the spectrum of the Dirac operator,
$\Dirac\phi_n=\lambda_n\phi_n$,\footnote{$\Dirac$ does not commute in
  general with $\Dirac^\dagger$ therefore it might not have a complete
  set of eigenvectors. This is no impediment for applying the
  $\zeta$-function method, which works for matrices of arbitrary
  Jordan form.}
\begin{equation}
\Tr(\Dirac^s)= \sum_n\lambda_n^s \,.
\end{equation}
Now, it follows that two Dirac operators related by a similarity
transformation $\Dirac^\prime=S^{-1} \Dirac S$ have the same spectrum,
and so the same effective action, within this regularization. This
proves that all classical symmetries which are realized by similarity
transformations can be preserved quantum-mechanically, if desired.
That is, anomalies in this class of symmetries are not essential, in
the sense that they can be removed by a suitable choice of
counter-terms.

$W$ can be split into normal and abnormal parity components,
\begin{equation}
W=W^+ + W^- \,.
\end{equation}
$W^\pm$ are the components which are even and odd, respectively, under
the pseudo-parity (or intrinsic parity) transformation: ${\cal S}\to
+{\cal S}$, ${\cal V}_\mu\to +{\cal V}_\mu$, ${\cal P}\to -{\cal P}$,
${\cal A}_\mu\to -{\cal A}_\mu$.  Equivalently, $W^+$ is the component
without Levi-Civita pseudo-tensor, is real (in Euclidean space) and
even under the exchange $R\leftrightarrow L$, while $W^-$ is the
component that contains the Levi-Civita pseudo-tensor, is purely
imaginary and odd under under the exchange of chiral labels
$R\leftrightarrow L$.  Since $W^\pm$ are the real and imaginary parts,
respectively, of $W$ one has, formally,
\begin{equation}
W^+= -\frac{1}{2}\Tr\,\log(\Dirac^\dagger\Dirac) 
\,,\qquad
W^- =  -\frac{1}{2}\Tr\,\log(\Dirac^{\dagger-1}\Dirac) 
\,.
\label{eq:2a.14}
\end{equation}
$W^+$ is theoretically better understood than $W^-$ which is more
challenging. Correspondingly, in this work our main focus will be on
the abnormal parity component.

In a derivative expansion of $W$, the terms are classified by the
number of covariant derivatives they carry. Due to Lorentz invariance,
for $d$ even there are only terms of even order (since the only
invariant tensors, the metric and the Levi-Civita pseudo-tensor, both
have an even number of indices to be contracted.) All terms with more
than $d$ derivatives in both $W^+$ and $W^-$ are UV finite.  The
expansion of $W^+$ starts at zero derivatives. The abnormal parity
component starts at order $d$, due to the presence of the Levi-Civita
pseudo-tensor. So in $W^-$ the leading order (LO)   is the only term
affected by UV ambiguities.

\subsection{Chiral symmetry}

The class of operators described in (\ref{eq:2.3}) is invariant under
the group of local chiral transformations. Let $\Omega_R(x)$ and
$\Omega_L(x)$ be matrices in internal space, assumed to be nowhere
singular. (In fact unitary, in order to preserve the hermiticity
properties of the external fields.) Then, the chirally rotated Dirac
operator is
\begin{equation}
\Dirac^\Omega= \thru{\D}{}^\Omega_R P_R \, +\thru{\D}{}^\Omega_L P_L 
+ \mlr^\Omega P_R+\mrl^\Omega P_L \,
\end{equation}
with
\begin{equation}
(v_\mu^{R,L})^\Omega= \Omega_{R,L}^{-1}v_\mu^{R,L} \Omega_{R,L} + 
\Omega_{R,L}^{-1}[\partial_\mu,\Omega_{R,L}]
\,,\qquad
\mlr^\Omega= \Omega_L^{-1}\mlr \Omega_R
\,,\qquad
\mrl^\Omega= \Omega_R^{-1}\mrl \Omega_L
\,.
\end{equation}
Also,  $(D_\mu^{R,L})^\Omega= \Omega_{R,L}^{-1} D_\mu^{R,L}\Omega_{R,L}$.

This is a symmetry of the classical theory. Indeed, the action of
$\Dirac$ with the configurations $\psi$ and $\bar\psi$, is the same as
that of $\Dirac^\Omega$ with rotated configurations
$\psi^\Omega=(\Omega_R^{-1}P_R+\Omega_L^{-1}P_L)\psi$, and
$\bar\psi^\Omega=\bar\psi(\Omega_L P_R+\Omega_R P_L)$.

Alternatively,  let us note that the Dirac operator can be written as
\begin{equation}
\Dirac=  P_L \thru{\D}_R P_R  +  P_R \thru{\D}_L P_L +   P_R \mlr P_R
+  P_L \mrl P_L \,,
\end{equation}
and so, in a convenient matrix form,
\begin{equation}
\Dirac= 
\left(
 \begin{matrix}
\mlr  &  \thru{\D}_L \cr
\thru{\D}_R   &   \mrl 
\end{matrix}
\right)
,
\label{eq:2a.18}
\end{equation}
the entries corresponding to the chiral subspaces $\gamma_5=\pm 1$.
Likewise
\begin{equation}
\Dirac^\Omega = 
\left(
 \begin{matrix}
\Omega_L^{-1}  &  0\cr
 0   & \Omega_R^{-1} 
\end{matrix}
\right)
\left(
 \begin{matrix}
\mlr  &  \thru{\D}_L \cr
\thru{\D}_R   &   \mrl 
\end{matrix}
\right)
\left(
 \begin{matrix}
\Omega_R  &  0\cr
 0   & \Omega_L
\end{matrix}
\right) .
\label{eq:2a.19}
\end{equation}
Then if $\psi= \left(
 \begin{matrix}
\psi_R \cr \psi_L 
\end{matrix}
\right)$ is a solution of $\Dirac\psi=0$, $\psi^\Omega=\left(
 \begin{matrix}
\Omega_R^{-1}  &  0\cr
 0   & \Omega_L^{-1}
\end{matrix}
\right)
 \left(
 \begin{matrix}
\psi_R \cr \psi_L 
\end{matrix}
\right)$ is a solution of $\Dirac^\Omega\psi^\Omega=0$.

As is well known classical symmetries may not survive
quantum-mechanically.  The property $\det AB=\det A \,\det B$, or $\tr
\log(AB)=\tr\log A + \tr\log B$ holds for matrices. This property
formally extends to operators, that is, it holds modulo UV
ambiguities. In the chiral case this implies for the effective
action 
\begin{equation}
W(\Dirac^\Omega)=W(\Dirac)+A(\Dirac,\Omega)
\end{equation}
where $A(\Dirac,\Omega)$ is an $\Omega$-dependent polynomial
counter-term (polynomial with respect to $\Dirac$) allowed by UV
ambiguity in the definition of $W$ (since $W(\Dirac^\Omega)$ and
$W(\Dirac)$ qualify both as valid determinations of the effective
action of $\Dirac$ owing to the classical symmetry property).

$A(\Dirac,\Omega)$ is a quantum-mechanical anomaly implying that $W$
is not chirally invariant under local transformations.  In the
literature, the name anomaly, or more precisely consistent anomaly,
refers to $A(\Dirac,\Omega)$ for infinitesimal $\Omega_{R,L}$.

Of course, part of the anomaly may come from a poor choice of
$W_{\text{ct}}$ in (\ref{eq:2a.9}). Actually, this is the case for
vector transformations.  These are the transformations of the
type $\Omega_R(x)=\Omega_L(x)$. This follows from our previous
observation that similarity transformations of $\Dirac$ do not change
the effective action if the $\zeta$-function prescription is adopted.
Vector transformations are similarity transformations,
$\Dirac^{\Omega_V}= \Omega_V^{-1}\Dirac\Omega_V$, therefore this
symmetry needs not be spoiled at the quantum-mechanical level.

Full chiral symmetry is not protected by this mechanism. Because chiral
rotations do not act as similarity transformations of $\Dirac$,
cf. (\ref{eq:2a.19}), the spectrum is not preserved and an anomaly is
introduced.  Nevertheless, the anomaly can be restricted to the
abnormal parity sector. Indeed, the adjoint Dirac operator
\begin{equation}
\Dirac^\dagger= 
\left(
 \begin{matrix}
\mrl  &  -\thru{\D}_R \cr
-\thru{\D}_L   &   \mlr 
\end{matrix}
\right)
\end{equation}
transforms as 
\begin{equation}
\Dirac^\Omega{}^\dagger = 
\left(
 \begin{matrix}
\Omega_R^{-1}  &  0 \cr
 0   & \Omega_L^{-1} 
\end{matrix}
\right)
\left(
 \begin{matrix}
\mrl  &  -\thru{\D}_R \cr
-\thru{\D}_L   &   \mlr 
\end{matrix}
\right)
\left(
 \begin{matrix}
\Omega_L  &  0 \cr
 0   & \Omega_R
\end{matrix}
\right) \,.
\label{eq:2a.22}
\end{equation}
Therefore $\Dirac^\dagger\Dirac$ does transform under a similarity
transformation and hence, from (\ref{eq:2a.14}), it follows that
$W^+(\Dirac)$ can be chosen to be chirally invariant. The remaining
anomaly in $W^-(\Dirac)$ cannot be completely removed. Its minimal
form (applying counter-terms to remove non essential contributions) is
the standard Bardeen's form \cite{Bardeen:1969md} if one chooses
vector transformations to be non anomalous. This is the VA
(vector-axial vector) form of the anomaly. Alternatively one can
choose the LR form in which the anomaly is composed of two terms, one
depending only on $v^R_\mu$ and $\Omega_R$ and another depending only
on $v^L_\mu$ and $\Omega_L$.  We choose the latter in this work and
this choice fully fixes the LR form of the effective action in the
abnormal parity sector.

For later reference, let us note that in the $\zeta$-function
regularization the axial anomaly (there is no vector anomaly) takes
the form $\Tr\left[ \gamma_5(\delta\Omega_L-\delta\Omega_R)
\Dirac^s\right]\big|_{s=0}$
\cite{Salcedo:1996qy}. Formally $\Dirac^s\big|_{s=0}$ is the identity
operator so there is a conflict of limits between the $\infty$ of the
trace of the multiplicative operator $\delta\Omega_L-\delta\Omega_R$
and the 0 from the trace of $\gamma_5$. A similar mechanism takes
place for the chiral anomaly in any renormalization scheme.

\subsection{WZW term and invariant remainder}
\label{subsec:2.D}

The variation of the effective action under a finite chiral
transformation can be obtained by integration of the infinitesimal
variation (the consistent anomaly). More specifically,
let $(\mlr,\mrl,v_R,v_L)$ be the field configuration obtained by applying the
chiral rotation $(\Omega_R,\Omega_L)$ to the configuration
$(\overline{m}_{LR},\overline{m}_{RL},\overline{v}_R,\overline{v}_L)$, then
\begin{equation}
W(m,v)-W(\overline{m},\overline{v})
=
\Gamma(v_R,\Omega_R)-\Gamma(v_L,\Omega_L)
,
\qquad
(v,m)=(\overline{m},\overline{v})^\Omega
,
\label{eq:2.23a}
\end{equation}
where $W$ refers to the LR effective action.  The function
$\Gamma(v,\Omega)$ verifies the obvious consistency condition
\begin{equation}
\Gamma(v,\Omega)= -\Gamma(\overline{v},\Omega^{-1})
.
\label{eq:2.24}
\end{equation}
Using $\Gamma(v,\Omega)$ a functional saturating the chiral anomaly
can be constructed, namely,\footnote{And conversely
  \cite{Salcedo:2000hx}, $\Gamma(v,\Omega) =
  \Gamma_{\text{WZW}}(v_R=v,v_L=0,U=\Omega) .  $}

\begin{equation}
\Gamma_{\text{WZW}}(v_R,v_L,U)=
\Gamma(v_R,\Omega_R)-\Gamma(v_L,\Omega_L) 
+
P_{\text{ct}}(\overline{v}_R,\overline{v}_L)
,
\qquad
U=\Omega_L^{-1}\Omega_R
.
\label{eq:b2}
\end{equation}
Here $P_{\text{ct}}(v_R,v_L)$ is a polynomial known as Bardeen's
subtraction \cite{Bardeen:1969md}. This is the counter-term needed to
pass from the LR form of the effective action to its VA form. E.g., in
two dimensions 
\begin{equation}
P_{\text{ct}}(v_R,v_L)= \frac{i}{4\pi}\int
d^2x\,\epsilon_{\mu\nu}v^R_\mu v^L_\nu
,
\qquad
(d=2)
.
\end{equation}
Of course, in any dimension
$P_{\text{ct}}(v_R,v_L)=\Gamma_{\text{WZW}}(v_R,v_L,U=1)$.

The (gauged) Wess-Zumino-Witten (WZW) functional
$\Gamma_{\text{WZW}}(v_R,v_L,U)$ depends on $v_{R,L}$ and a field $U$
and saturates the anomaly by construction provided only that $U$
transforms as $U\to\Omega_L^{-1}U\Omega_R$. (Note that although the
WZW term depends on $U$, this dependence cancels in the anomaly.)  For
instance, in two dimensions
\begin{eqnarray}
\Gamma_{\text{WZW}}(v_R,v_L,U) &=&
-\frac{i}{12\pi}\int\epsilon_{\mu\nu\alpha}
\tr\big(
U^{-1}\partial_\mu U\,
 U^{-1}\partial_\nu U \,
U^{-1}\partial_\alpha U
\big)
\, d^3x
\nonumber \\
&&
+\frac{i}{4\pi}\int\epsilon_{\mu\nu}
\tr\big(
-\partial_\mu U \, U^{-1} v^L_\nu
-U^{-1}\,\partial_\mu U  \, v^R_\nu
+ U \,v^R_\mu \,U^{-1} v^L_\nu
\big)
\, d^2x
.
\end{eqnarray}

Since $\mlr$ and $\mrl^{-1}$ both transform as required for $U$, the
anomaly is saturated by the following effective action
\begin{eqnarray}
\Gamma_{\text{gWZW}} 
&=&
\frac{1}{2}\left(
\Gamma_{\text{WZW}}(v_R,v_L,\mlr)
+\Gamma_{\text{WZW}}(v_R,v_L,\mrl^{-1})
\right)
\nonumber \\
&=&
\frac{1}{2}\left(
\Gamma_{\text{WZW}}(v_R,v_L,\mlr)
-\Gamma_{\text{WZW}}(v_L,v_R,\mrl)
\right)
,
\label{eq:b1}
\end{eqnarray}
which is odd under the exchange $L\leftrightarrow R$. That is,
\begin{equation}
  \Gamma_{\text{gWZW}}(m,v)-\Gamma_{\text{gWZW}}(\overline{m},\overline{v})
  =
  \Gamma(v_R,\Omega_R)-\Gamma(v_L,\Omega_L)
  ,
  \qquad
  (v,m)=(\overline{m},\overline{v})^\Omega
  .
\end{equation}
Explicitly, in
two dimensions,
\begin{eqnarray}
\Gamma_{\text{gWZW}} &=&
-\frac{i}{24\pi}\int\epsilon_{\mu\nu\alpha}
\tr\big(
\mlr^{-1}\partial_\mu\mlr \,
 \mlr^{-1}\partial_\nu\mlr \,
\mlr^{-1}\partial_\alpha\mlr 
-
\mrl^{-1}\partial_\mu\mrl \,
 \mrl^{-1}\partial_\nu\mrl \,
\mrl^{-1}\partial_\alpha\mrl
\big)
\, d^3x
\nonumber \\
&&
+\frac{i}{8\pi}\int\epsilon_{\mu\nu}
\tr\big(
\partial_\mu\mrl \, \mrl^{-1} v^R_\nu
-\partial_\mu\mlr \, \mlr^{-1} v^L_\nu
-\mlr^{-1}\partial_\mu\mlr  \, v^R_\nu
+\mrl^{-1}\partial_\mu\mrl  \, v^L_\nu
\nonumber \\
&& \qquad\quad
-\mrl \,v^L_\mu \mrl^{-1} v^R_\nu
+\mlr \,v^R_\mu \mlr^{-1} v^L_\nu
\big)
\, d^2x
.
\label{eq:2.25}
\end{eqnarray}
Actually this is a generalized Wess-Zumino-Witten term since
$\mrl,\mlr$ are not restricted to lie on the chiral
circle.\footnote{The chiral circle constraint corresponds to $\mlr= M
  U$, $\mrl= M U^{-1}$, where $U$ is unitary and $M$ is a constituent
  mass that cancels in $\Gamma_{\text{gWZW}}$.}

Therefore one can write
\begin{equation}
W^- =   W^-_c + \Gamma_{\text{gWZW}}
,
\label{eq:2.23}
\end{equation}
where $W^-_c$ is chirally invariant and $\Gamma_{\text{gWZW}}$
reproduces the anomaly. Of course, one could transfer contributions
from $W^-_c$ to $\Gamma_{\text{gWZW}}$ and the latter would still
saturate the anomaly, however, our choice of $\Gamma_{\text{gWZW}}$ is
distinguished in the sense that it is composed of terms depending only
on $\mlr$ plus terms depending only on $\mrl$ and contains just LO
terms in the derivative expansion.

We will refer to $W_c^-$ as the chiral remainder, that is, the chiral
invariant terms left after the anomaly saturating part
$\Gamma_{\text{gWZW}}$ has been subtracted. At leading order in a
derivative expansion, the remainder $W^-_c$ vanishes identically when
the scalar and pseudo-scalar fields satisfy a generalized chiral
circle constraint (namely, when $\mrl\mlr$ is a c-number) but $W^-_c$
is a non trivial functional outside the chiral circle
\cite{Salcedo:2000hx} or beyond LO \cite{Hernandez:2007ng}.

\subsection{Computation of $W^-$ from the current}

The operator $\Dirac^\dagger\Dirac$ is of the Klein-Gordon type
therefore there are several techniques to address the computation of
$W^+$. This is further simplified by the fact that chiral symmetry is
preserved in the normal parity sector. This helps to reduce the number
of allowed structures.

The situation in the abnormal parity sector is quite different since
chiral symmetry is not preserved there. This means that $W^-$ cannot
be written using simple chiral covariant blocks like
$F^{R,L}_{\mu\nu}$, $\mrl$, $\mlr$ and their chiral covariant
derivatives and this complicates considerably its calculation and even
its proper mathematical definition beyond perturbation theory
\cite{Ball:1989xg}.  In addition, the operator
$\Dirac^{\dagger-1}\Dirac$ in (\ref{eq:2a.14}) is not of the
Klein-Gordon type, in fact, it is not even local. The obvious method
is to use $\Dirac^2$ as Klein-Gordon operator (to obtain $W=
-\frac{1}{2}\Tr\log(\Dirac^2)$).  The computation through the heat
kernel of $\Dirac^2$ or its $\zeta$-function becomes quite involved
due to the lack of full chiral symmetry (vector symmetry is
preserved).

Since exploiting full chiral symmetry is essential, the route to
attack the problem has been to use the current. The current is defined
as the variation of the effective action with respect to the gauge
fields:
\begin{equation}
\delta_v W= \int d^dx \,
\tr(J^R_\mu(x)\,\delta v^R_\mu(x)+J^L_\mu(x)\,\delta v^L_\mu(x)) .
\end{equation}
Here the trace includes only internal degrees of freedom (not Dirac
ones).  Formally,
\begin{equation}
\delta_v W = -\Tr\left(\frac{1}{\Dirac}\,\delta_v \Dirac\right),
\qquad
\delta_v \Dirac= \delta\!\thrur{v}_R \!P_R +\delta\!\thrur{v}_L\! P_L .
\end{equation}
Although this quantity is more UV convergent than the effective
action, a direct computation of the current using this expression is
still subject to UV ambiguities and different determinations of the
current differ by local polynomial counter-terms of mass dimension
$d-1$. Unlike $W^-$, and this is the key point of using the current to
compute the effective action, the current in the abnormal parity
sector can be computed preserving local chiral covariance. This is
known as the covariant version of the current, $J^{R,L}_{c,\mu}$.
Unfortunately, as a consequence of the chiral anomaly, the covariant
current (of the abnormal parity sector) is not consistent, that is, it
is not a true variation of any effective action. The consistent
current is obtained by adding the appropriate counter-term
\cite{Bardeen:1984pm}:
\begin{equation}
J^{R,L}_\mu= J^{R,L}_{c,\mu} + P^{R,L}_\mu
\qquad\text{($W^-$ sector)}
.
\end{equation}
The current counter-term is fully fixed by the chiral anomaly (once,
e.g., the LR version has been adopted). For instance, for $d=2$, one
finds $P^{R,L}_\mu= i\epsilon_{\mu\nu}v^{R,L}_\nu/4\pi$. This term is
just the polynomial non covariant part of the current derived from
$\Gamma_{\text{gWZW}}$ in (\ref{eq:2.25}).

Let us remark that the current
coming from $W^-_c$ is both covariant and consistent, and does not
coincide with the covariant current $J_{c,\mu}^{R,L}$, which is not
consistent. The covariant current picks up some covariant terms from
$\Gamma_{\text{gWZW}}$.

Once the consistent current is obtained the effective action can be
reconstructed from it. This method is the basis of \cite{Ball:1989xg}
to achieve a suitable definition of $W^-$ beyond perturbation theory.

The approach based on the current has the virtue of being fully chiral
covariant. It has been used in the derivative expansion calculations
of \cite{Salcedo:2000hx} and \cite{Hernandez:2007ng}, where the
unknown $W_c^-$ is determined so that when added to the known WZW term
the consistent current is reproduced.

\section{The invariant part of the effective action}
\label{sec:3a}

In this section the main result of the paper is presented, namely, we
show how, upon separation of the anomalous WZW contribution, the
effective action can also be expressed as the $\Tr\log$ of a local
operator of the Klein-Gordon type which, in addition, is manifestly
chiral covariant.

\subsection{Covariant Klein-Gordon operator}

In order to construct such a covariant operator of the Klein-Gordon
type, we will use the convenient matrix notation of (\ref{eq:2a.18}).
That is, provided that the operators $A$ and $D$ commute with
$\gamma_5$ and $B$ and $C$ anticommute with $\gamma_5$, we can use
$\left(
 \begin{matrix}
A  &  B \cr
C   & D
\end{matrix}
\right) $ to represent $P_R A P_R + P_R B P_L + P_L C P_R + P_L D
P_L$.  These matrices multiply as usual. The only thing to be noted is
the relation
\begin{equation}
\Tr \left(
 \begin{matrix}
A  &  B \cr
C   & D
\end{matrix}
\right)
=\frac{1}{2}\Tr(A+D)+\frac{1}{2}\Tr(\gamma_5(A-D))
.
\label{eq:2a.26}
\end{equation}

In this notation
\begin{eqnarray}
\Dirac= 
\left(
 \begin{matrix}
\mlr  &  \thru{\D}_L \cr
\thru{\D}_R   &   \mrl 
\end{matrix}
\right)
.
\end{eqnarray}
Let us introduce the two new operators
\begin{eqnarray}
\Dirac^\prime &=&
\left(
 \begin{matrix}
\mlr  &  -\thru{\D}_L \cr
-\thru{\D}_R   &   \mrl 
\end{matrix}
\right)
\nonumber \\
\Dirac^m &=&
\left(
 \begin{matrix}
\mlr^{-1}  & 0 \cr
0   &   \mrl^{-1} 
\end{matrix}
\right)
\Dirac
\left(
 \begin{matrix}
\mrl  &  0 \cr
0   &   \mlr 
\end{matrix}
\right)
=
\left(
 \begin{matrix}
\mrl  &  \mlr^{-1}\thru{\D}_L\mlr \cr
\mrl^{-1}\thru{\D}_R\mrl  &   \mlr 
\end{matrix}
\right)
,
\label{eq:3.3}
\end{eqnarray}
and their product
\begin{eqnarray}
\Kop &=&
\Dirac^\prime
\Dirac^m 
=
\left(
 \begin{matrix}
K_L  & 0 \cr
0   &   K_R 
\end{matrix}
\right)
,
\nonumber \\
K_R &=& \mrl\mlr - \thru{\D}_R  \mlr^{-1} \thru{\D}_L \mlr
,
\qquad
K_L=\mlr\mrl - \thru{\D}_L \mrl^{-1} \thru{\D}_R \mrl
.
\label{eq:main}
\end{eqnarray}
These operators are related by $K^\dagger_R= \mrl K_L\mrl^{-1}$ and
$K^\dagger_L= \mlr K_R \mlr^{-1}$.\footnote{
It might look awkward that $K_L$ (a
  chirally left operator) appears in the subspace $\gamma_5=+1$
  instead of $\gamma_5=-1$ (and similarly for $K_R$).  The natural
  assignments would be achieved by using instead 
$
\Dirac^m{}^\prime =
\left(
 \begin{matrix}
\mrl  &  0 \cr
0   &   \mlr 
\end{matrix}
\right)
\Dirac
\left(
 \begin{matrix}
\mlr^{-1}  & 0 \cr
0   &   \mrl^{-1} 
\end{matrix}
\right)
$ and $\Kop^\prime=\Dirac^m{}^\prime\Dirac^\prime
=
\left(
 \begin{matrix}
K_R^\dagger  & 0 \cr
0   &   K_L^\dagger 
\end{matrix}
\right) $.  Both treatments are equivalent (the same inversion takes
place if one uses $\Tr\log(\Dirac\Dirac^\dagger)$ instead of
$\Tr\log(\Dirac^\dagger\Dirac)$). We favor $K_{R,L}$ since we prefer
the structure $U^{-1} d U$ over $dU U^{-1}$. The same effect would be
obtained by changing $\gamma_5\to-\gamma_5$, so that (\ref{eq:2.3})
becomes $\Dirac= P_R \thru{\D}_R +P_L \thru{\D}_L + P_L \mlr + P_L
\mrl$.  }

Chirally, $\Dirac^\prime$ transforms as $\Dirac$ while $\Dirac^m$
transforms as $\Dirac^\dagger$.  As a consequence $\Kop$ is a
manifestly covariant operator (i.e., chiral rotations act as
similarity transformations on it) which is local and of the
Klein-Gordon type.  This is the operator we were looking for.  A
formal hand-waving argument shows that $(\Det\Kop)^{1/2}$ is the
chiral invariant part of the chiral determinant $\Det\Dirac$.  This is
as follows:
\begin{eqnarray}
W &=& -\Tr\,\log \Dirac
=
-\frac{1}{2}\Tr\,\log \Dirac^\prime\Dirac
\nonumber \\
&=&
-\frac{1}{2}\Tr\,\log 
\Dirac^\prime\Dirac^m
\left(
 \begin{matrix}
\mrl^{-1}\mlr  & 0 \cr
0   &    \mlr^{-1}\mrl
\end{matrix}
\right)
\nonumber \\
&=&
-\frac{1}{2}\Tr\,\log
\Kop
+
\frac{1}{2}\Tr\left(\gamma_5\log(\mlr^{-1}\mrl)\right)
.
\label{eq:3.5}
\end{eqnarray}
In the second equality we have used that $\Dirac$ and $\Dirac^\prime$
have the same effective action. This is because the trace of products
involving an odd number of $\gamma_\mu$ (with or without $\gamma_5$)
vanish, and so only terms with an even number of $\thru{\D}_{R,L}$
will give a contribution to $W$. (In other way, in even dimensions the
representations $\gamma_\mu$ and $-\gamma_\mu$ are equivalent.)
Therefore, we can symmetrize with respect to $\thru{\D}_{R,L}\to
-\thru{\D}_{R,L}$. In the third and fourth equalities we make use of
the formal identity
\begin{equation}
\Tr \log(AB)=\Tr\log A + \Tr\log B
.
\end{equation}
It implies that (formally) operators commute inside $\Tr\,\log$ and so
the factors can be rearranged at will.

In the right-hand side of (\ref{eq:3.5}) $-\frac{1}{2}\Tr\,\log \Kop$
is a chiral invariant contribution whereas the second term
$\frac{1}{2}\Tr\left(\gamma_5\log(\mlr^{-1}\mrl)\right)$ represents
the anomalous part. This latter term has three conspicuous features:
first, just as the anomaly, it presents an indetermination of the type
$0\,\infty$ ($0$ from $\tr\gamma_5=0$ and $\infty$ from the trace of
the multiplicative operator $\mlr^{-1}\mrl$). This is typical of
anomalous contributions.  Second, although
$\Tr\left(\gamma_5\log(\mlr^{-1}\mrl)\right)$ is not chiral invariant,
its variation is invariant. Indeed,
\begin{equation}
\delta\Tr\left(\gamma_5\log(\mlr^{-1}\mrl)\right)
=
-\Tr\left(\gamma_5 \mlr^{-1}\delta\mlr  \right)
+\Tr\left(\gamma_5 \mrl^{-1}\delta\mrl  \right)
.
\end{equation}
This is also characteristic of the anomalous WZW term: its variation
is covariant up to polynomial counter-terms (the latter are missed by
our formal manipulations). Finally, a third feature is that this last
term is separable in two contributions, one depending only on $\mlr$
and one depending only on $\mrl$, just as $\Gamma_{\text{gWZW}}$.

In view of this, comparison of (\ref{eq:3.5}) with (\ref{eq:2.23})
suggests identifying the covariant part with $W^++W^-_c$, while the non
covariant part would represent the anomalous contribution,
$\Gamma_{\text{gWZW}}$.  The latter cannot be recovered from the
formal expression. That is,
\begin{equation}
W = 
-\frac{1}{2}\Tr\,\log
\Kop
+
\Gamma_{\text{gWZW}}
\label{eq:2.36}
,
\end{equation}
or
\begin{equation}
W_c(\Dirac) = 
-\frac{1}{2}\Tr\,\log
\Kop
.
\end{equation}

It will be important to realize that within the derivative expansion
all orders beyond the LO are UV convergent.  This LO contains the UV
ambiguities and because different orders are not mixed by chiral
rotations, it carries all the anomalous contributions. This implies
that the formal manipulations used to arrive to (\ref{eq:3.5}) are
correct beyond LO in the abnormal parity sector and so (\ref{eq:2.36})
certainly holds to all orders beyond the lowest one. In fact we would
expect it to hold to all orders, included the LO one.  The reason is
that in most expansions of interest (such as perturbation theory or
inverse mass expansions) higher orders are increasingly UV convergent
and so (\ref{eq:2.36}) should be fulfilled to all UV convergent orders
of all those expansions. This covers many contributions which belong
to the LO from the point of view of the derivative expansion. This
expectation is indeed correct. As we show subsequently, the relation
(\ref{eq:2.36}) holds for all (even) space-time dimensions.

\subsection{Proof  of the main result}

In order to prove the identification in (\ref{eq:2.36}) we should make
the statement precise.

\subsubsection{The standard LR effective action of $\Dirac$}

The functional $W^-(\Dirac)$ is perfectly well defined once the two
following conditions are met, first, the LR version of the anomaly is
chosen, and second, $W^-(\Dirac)$ depends only on $\Dirac$ (no new
field absent from $\Dirac$ is introduced in the functional).  This is
unique because the allowed ambiguity would be an abnormal parity and
chiral invariant polynomial composed of $F^{R,L}_{\mu\nu}$ and
$\mlr,\mrl$ and their covariant derivatives, but no such polynomial
exists. (We disregard topological contributions, such as $\int
d^2x\epsilon_{\mu\nu}\tr(F^R_{\mu\nu}-F^L_{\mu\nu})$ in two
dimensions.) Since $\Gamma_{\text{gWZW}}$ is also well defined,
$W^-_c(\Dirac)$ is unique too. On the other hand $W^+(\Dirac)$ is not
unique. A basic definition would be, e.g., its $\zeta$-function
determination from $\Dirac^\dagger\Dirac$. New chiral invariant
determinations are then obtained by adding arbitrary normal parity and
chiral invariant polynomials of $F^{R,L}_{\mu\nu}$ and $\mlr,\mrl$ and
their covariant derivatives. These do not vanish identically, e.g.,
$\int d^2x\,\tr(\mrl\mlr)$ in two dimensions. For concreteness we
assume that the above mentioned basic determination has been taken for
$W^+(\Dirac)$. We will use the notation $W_s(\Dirac)$ (with subindex
$s$ from ``standard'') to denote this specific determination of the
effective action:
\begin{equation}
W_s(\Dirac)= W^-_{\text{LR}}(\Dirac)-\frac{1}{2}\frac{d}{ds}\Tr((\Dirac^\dagger\Dirac)^s)\Big|_{s=0}
.
\end{equation}

\subsubsection{Natural and standard effective action from $\Kop$}

Let us now consider the functional
\begin{equation}
W(\Kop)= -\frac{1}{2}\Tr\log\Kop \,.
\end{equation}
As noted before, given a differential operator, such as $\Dirac$ or
$\Kop$, the logarithm of its determinant is unique modulo a
counter-term action which is polynomial regarding its dependence on
the fields present in the operator and their derivatives. A basic
definition of $W(\Kop)$ follows from its $\zeta$-function
determination. Two nice properties of this determination are i) it
does not introduce new fields in the game, and ii) it preserves all
symmetries realized as similarity transformations of $\Kop$ (including
chiral symmetry). We will refer to determinations with these two
properties as ``natural'' determinations. All determinations of the
type $\Tr f(\Kop,s)$ with $f(x,s)\to\log(x)$ as $s\to 0$, are of
natural type. All natural determinations differ from the
$\zeta$-function one by a chiral invariant polynomial constructed with
the fields in $\Kop$ and their derivatives. We will take the
$\zeta$-function determination as the standard one, to be denoted
$W_s(\Kop)$:
\begin{equation}
W_s(\Kop)= -\frac{1}{2}\frac{d}{ds}\Tr(\Kop^s)\Big|_{s=0}
.
\end{equation}

It order to analyze this point further, let us
introduce the chiral covariant quantities
\begin{eqnarray}
M_R=\mrl\mlr \,,\quad 
M_L=\mlr\mrl \,,
\nonumber \\
\R^R_\mu= \mlr^{-1}(D_\mu m)_{LR} \,,
\quad
\R^L_\mu= \mrl^{-1}(D_\mu m)_{RL} \,,
\end{eqnarray}
where
\begin{equation}
(D_\mu m)_{LR} = D^L_\mu\mlr-\mlr D^R_\mu\,,
\quad
(D_\mu m)_{RL} = D^R_\mu\mrl-\mrl D^L_\mu \,.
\end{equation}
In terms of these fields
\begin{eqnarray}
\Dirac^m &=&
\left(
 \begin{matrix}
\mrl  &  \thru{\D}_R+\thru{\R}_R \cr
 \thru{\D}_L+\thru{\R}_L  &   \mlr 
\end{matrix}
\right)
,
\label{eq:3.14a}
\end{eqnarray}
and
\begin{eqnarray}
\Kop &=&
\left(
 \begin{matrix}
M_L \,-\thru{\D}_L^{\,\,2}  -\thru{D}_L \,\thru{\R}_L
  & 0 \cr
0   &
M_R \, -\thru{\D}_R^{\,\,2}  -\thru{D}_R \,{\!\thru{\R}}_R
\end{matrix}
\right)
.
\end{eqnarray}
Therefore, within the class of natural determinations, the ambiguity
in $W(\Kop)$ is a polynomial in $M_{L,R}$, $D^{L,R}_\mu$, and
$\R^{L,R}_\mu$. Unfortunately, $\R^{R,L}_\mu$ is not a polynomial with
respect the original fields in $\Dirac$. This implies that, in
general, an UV ambiguity that would be admissible from the point of
view of $W(\Kop)$, will not be admissible for $W(\Dirac)$, when
inserted in (\ref{eq:2.36}). That is, even if $W(\Kop)$ is obtained
through a natural determination, one must still allow for removal by
counter-terms of contributions which are not polynomials with respect
to $\Dirac$ (and so are incorrect) but polynomials with respect to
$\Kop$. Such polynomials are absent in two dimensions in the abnormal
parity sector.\footnote{The possible candidates, $\int
  d^2x\epsilon_{\mu\nu}\tr(\R^R_\mu \R^R_\nu-\R^L_\mu \R^L_\nu) $, and
  $\int d^2x\epsilon_{\mu\nu}\tr([D^R_\mu,\R^R_\nu]-[D^L_\mu,\R^L_\nu])
  $, vanish.  } But they exists in four or more dimensions. They also
exist in two dimensions in the normal parity sector, e.g., $\int d^2
x\,\tr(\R^R_\mu \R^R_\mu+\R^L_\mu \R^L_\mu) $.

\subsubsection{Proof of the statement}

The statement to be proven is then that for a certain natural
determination of $W(\Kop)$ the relation (\ref{eq:2.36}) holds. Or
equivalently,
\begin{eqnarray}
W_s(\Dirac)-\Gamma_{\text{gWZW}}(\Dirac)
-W_s(\Kop)
& &
\nonumber \\
&&
\hskip -3.5cm
= \text{``Chiral invariant polynomial of $\mlr,\mrl,D^{L,R}_\mu$, and
$\R^{L,R}_\mu$''}
.
\label{eq:3a.17}
\end{eqnarray}
Note that in the abnormal parity sector the polynomial will depend
only on $D^{L,R}_\mu$, and $\R^{L,R}_\mu$, but not explicitly on
$\mlr,\mrl$ due to dimensional reasons: there should be precisely $d$
fields carrying a Lorentz index and they already saturate the mass
dimension $d$ of the counter-term.  (As always, the polynomial has
degree at most $d$ and this holds too for similar polynomials below.)

As we argued before $W_s(\Dirac^\prime)=W_s(\Dirac)$. On the other
hand, from the relation $\Kop=\Dirac^\prime\Dirac^m$ we can conclude
that $\Tr\log\Kop$ can be computed through
$\Tr\log\Dirac^\prime+\Tr\log\Dirac^m$, modulo UV ambiguities, that is
\begin{eqnarray}
\frac{1}{2}(W_s(\Dirac)+W_s(\Dirac^m))
-
W_s(\Kop)
&&
\nonumber \\
&&
\hskip -3.5cm
= \text{``Polynomial of $\mlr,\mrl,\partial_\mu,v^{L,R}_\mu$, and
$\R^{L,R}_\mu$''}
.
\label{eq:3.18}
\end{eqnarray}
In this polynomial we have to allow for the variables $\R^{L,R}_\mu$,
since they appear in $\Dirac^m$, cf. (\ref{eq:3.14a}). Indeed, if we
represent the Dirac operator as
\begin{equation}
\Dirac=\Dirac(\mlr,\mrl,v_R,v_L)
\end{equation}
then
\begin{equation}
\Dirac^m= \Dirac\Big|_{v\to v+\R,  R\leftrightarrow L}
= \Dirac(\mrl,\mlr,v_L+\R_L,v_R+\R_R)
.
\end{equation}

From its definition, (\ref{eq:3.3}), it is clear that $\Dirac^m$ is
related to $\Dirac$ by a (generalized) chiral rotation (cf.
(\ref{eq:2a.19})) with $\Omega_R=\mrl$ and $\Omega_L=\mlr$. Therefore,
using (\ref{eq:2.23a}) and (\ref{eq:2.24}),
\begin{equation}
W_s(\Dirac^m)
=
W_s(\Dirac)-\Gamma(v_R,\mrl^{-1})+\Gamma(v_L,\mlr^{-1})
.
\end{equation}
Substituting in (\ref{eq:3.18}) we find
\begin{eqnarray}
W_s(\Dirac)
-
W_s(\Kop)
-\frac{1}{2}(
\Gamma(v_R,\mrl^{-1})-\Gamma(v_L,\mlr^{-1})
)
&&
\nonumber \\
&&
\hskip -3.5cm
= \text{``Polynomial of $\mlr,\mrl,\partial_\mu,v^{L,R}_\mu$, and
$\R^{L,R}_\mu$''}
.
\label{eq:3.18a}
\end{eqnarray}
This can be rewritten 
as
\begin{eqnarray}
W_s(\Dirac)-\Gamma_{\text{gWZW}}(\Dirac)
-W_s(\Kop)
&&
\nonumber \\
&&
\hskip -3.5cm
= 
P(\Dirac)+
\text{``Polynomial of $\mlr,\mrl,\partial_\mu,v^{L,R}_\mu$, and
$\R^{L,R}_\mu$''}
.
\label{eq:3.18b}
\end{eqnarray}
where we have defined
\begin{equation}
P(\Dirac)
=
\frac{1}{2}(
\Gamma(v_R,\mrl^{-1})-\Gamma(v_L,\mlr^{-1})
)
-
\Gamma_{\text{gWZW}}(\Dirac)
.
\end{equation}
Clearly, (\ref{eq:3.18b}) will be equivalent to (\ref{eq:3a.17})
provided that $P(\Dirac)$ is also a polynomial. This is the case as we
now show: using (\ref{eq:b1}) and (\ref{eq:b2})
\begin{eqnarray}
\Gamma_{\text{gWZW}}(\Dirac)
&=&
\frac{1}{2}\left[
\Gamma_{\text{WZW}}(v_R,v_L,\mlr)
-\Gamma_{\text{WZW}}(v_L,v_R,\mrl)
\right]
\nonumber \\
&=&
\frac{1}{2}\left[
\left(
\Gamma(v_R,\Omega_R)-\Gamma(v_L,\Omega_L) 
+
P_{\text{ct}}(v_R^{\Omega^{-1}_R},v_L^{\Omega^{-1}_L})
\right)\Big|_{\Omega_R=1,\Omega_L=\mlr^{-1}}
\right.
\nonumber \\
&&
-
\left.
\left(
\Gamma(v_L,\Omega_R)-\Gamma(v_R,\Omega_L) 
+
P_{\text{ct}}(v_L^{\Omega^{-1}_R},v_R^{\Omega^{-1}_L})
\right)\Big|_{\Omega_R=1,\Omega_L=\mrl^{-1}}
\right]
\nonumber \\
&=&
\frac{1}{2}\left[
\Gamma(v_R,\mrl^{-1})-\Gamma(v_L,\mlr^{-1})
+
P_{\text{ct}}(v_R,v_L^{\mlr})
-
P_{\text{ct}}(v_L,v_R^{\mrl})
\right]
.
\end{eqnarray}
In this expression
\begin{equation}
(v_L^{\mlr})_\mu
=
(v_L^\Omega)_\mu\Big|_{\Omega=\mlr}
=
(\Omega^{-1}\partial_\mu\Omega
+
\Omega^{-1}v^L_\mu\Omega
)\Big|_{\Omega=\mlr}
=
v^R_\mu+\R^R_\mu
,
\qquad
(v_R^{\mrl})_\mu
=
v^L_\mu+\R^L_\mu
.
\end{equation}
This implies
\begin{eqnarray}
P(\Dirac)
&=&
-\frac{1}{2}\left[
P_{\text{ct}}(v_R,v_R+Q_R)
-
P_{\text{ct}}(v_L,v_L+Q_L)
\right]
.
\end{eqnarray}
This is a polynomial constructed with $v^{R,L}_\mu$ and $\R^{R,L}_\mu$
as advertised. Because the left-hand side of (\ref{eq:3.18b}) is
chiral invariant, so is the polynomial on the right. This proves
(\ref{eq:3a.17}).

\subsection{Explicit Klein-Gordon form}

The operators $K_{R,L}$ can be brought to a manifest Klein-Gordon form
if desired. Indeed,
\begin{eqnarray}
K_R
&=&
\mrl\mlr - \thru{\D}_R  \mlr^{-1} \thru{\D}_L \mlr
\nonumber \\
&=&
\mrl\mlr - \thru{\D}_R^{\,\,2} 
-\thru{\D}_R\mlr^{-1}(\thru{\D}_L\mlr-\mlr\thru{\D}_R)
\nonumber \\
&=&
\mrl\mlr - (D^R_\mu)^2 
-\frac{1}{2}\sigma_{\mu\nu}F^R_{\mu\nu}
-D^R_\mu \mlr^{-1} (D_\nu m)_{LR} \gamma_\mu\gamma_\nu 
\nonumber \\
&=&
{\tilde M}_R - ({\tilde D}^R_\mu)^2 
\label{eq:4.9}
\end{eqnarray}
with
\begin{eqnarray}
{\tilde M}_R &=& \mrl\mlr
-\frac{1}{2}\sigma_{\mu\nu} F^R_{\mu\nu}
+ (B^R_\mu)^2 -[D^R_\mu, B^R_\mu] \,,
\nonumber \\
{\tilde D}^R_\mu &=& D^R_\mu+B^R_\mu\,,
\nonumber \\
B^R_\mu &=& \frac{1}{2}\gamma_\mu\gamma_\nu \,\mlr^{-1}(D_\nu m)_{LR}
=
\frac{1}{2}\gamma_\mu\gamma_\nu \,\R^R_\nu
\,.
\label{eq:4.10}
\end{eqnarray}
In these expressions we have used the relations
\begin{eqnarray}
\gamma_\mu\gamma_\nu=\delta_{\mu\nu}+\sigma_{\mu\nu},
\quad
F^R_{\mu\nu}=[D^R_\mu,D^R_\nu], 
\quad
(D_\mu m)_{LR}= D^L_\mu\mlr-\mlr D^R_\mu
.
\end{eqnarray}
Of course, there is an entirely analogous expression for $K_L$. This
gives (using (\ref{eq:2a.26}))
\begin{eqnarray}
  W^+
&=& 
-\frac{1}{4}\Tr\log({\tilde M}_R - ({\tilde D}^R_\mu)^2 ) 
  -\frac{1}{4}\Tr\log({\tilde M}_L - ({\tilde D}^L_\mu)^2 )
,
\nonumber \\
  W^-_c 
&=& 
\frac{1}{4}\Tr\left(\gamma_5\log({\tilde M}_R - ({\tilde D}^R_\mu)^2 ) \right)
  -\frac{1}{4}\Tr\left(\gamma_5\log({\tilde M}_L - ({\tilde D}^L_\mu)^2 ) 
\right)
.
\label{eq:2.36a}
\end{eqnarray}
Note that ${\tilde M}_{R,L}$ and ${\tilde D}_\mu^{R,L}$ commute with
$\gamma_5$. The second relation is quite remarkable: it allows to treat the
(chiral invariant part of the) imaginary part of the fermionic
effective action, $W^-$, with the same techniques available to attack
the real part, e.g., the heat kernel expansion, preserving manifest
chiral invariance throughout without any obstruction from the chiral
anomaly.

\subsection{Direct computation of the leading order in two dimensions}

In this section we carry out explicitly the computation of $W_c^-$ in
two dimensions to LO in the derivative expansion. This verifies the
identification (\ref{eq:2.36}) in this case and illustrates its use.

To this end we apply the convenient method of Chan \cite{Chan:1986jq}
to compute the trace of the logarithm of $K_R$. This takes the form
\begin{equation}
\Tr\left(\gamma_5\log({\tilde M}_R - ({\tilde D}^R_\mu)^2 ) \right)
= \int\frac{d^dx \, d^dp}{(2\pi)^d}
\tr\gamma_5\left(-\log{\tilde N}_R +
\frac{p^2}{d} [{\tilde D}^R_\mu, {\tilde N}_R]^2 + \cdots
\right) ,
\label{eq:3.14}
\end{equation}
where ${\tilde N}_R=1/(p^2+{\tilde M}_R)$ and the dots refer to higher orders
in the derivative expansion with respect to ${\tilde D}^R_\mu$.

Actually we want to expand with respect to the original covariant
derivatives $D^{R,L}_\mu$ (rather than ${\tilde D}^R_\mu$).  Noting
that $B^R_\mu$ is of first order (contains exactly one covariant
derivative) and ${\tilde M}_R$ is $\mrl\mlr$ plus terms of second
order, it follows that each given order in the derivative expansion of
$ \Tr(\gamma_5\log K_R)$ gets contributions from a finite number of
terms of the Chan expansion, therefore the computation is feasible
order by order using this scheme.

To proceed to LO in two dimensions requires the second order (two
derivative) contributions from the first two Chan terms $-\log{\tilde
  N_R}$ and $[{\tilde D}^R_\mu, {\tilde N}_R]^2$. Introducing
\begin{equation}
N_R= (p^2+\mrl\mlr)^{-1}
\end{equation}
one finds
\begin{eqnarray}
-\tr\left(\gamma_5\log({\tilde N}_R)\right)
&=&
 -\tr(\gamma_5\log N_R) 
+\tr\!\left(
\gamma_5 \,N_R\left(
-\frac{1}{2}\sigma_{\mu\nu} F^R_{\mu\nu}
+ (B^R_\mu)^2 -[D^R_\mu, B^R_\mu]
\right)\right)
+{\mathcal O}(D^4)
,
\nonumber \\
{}[{\tilde D}^R_\mu, {\tilde N}_R]^2
&=&
[D^R_\mu+B^R_\mu,N_R]^2
+{\mathcal O}(D^4)
.
\end{eqnarray}

Introducing these expressions in Chan's formula (\ref{eq:3.14}) and 
taking the Dirac trace using the two dimensional identity
$\tr(\gamma_5\gamma_\mu\gamma_\nu)=-2i\epsilon_{\mu\nu}$, gives
\begin{eqnarray}
\Tr(\gamma_5\log K_R)_{d=2,\text{LO}} &=&
2i\int \frac{d^2x d^dp}{(2\pi)^d}\epsilon_{\mu\nu}\tr\!\left(
\frac{1}{2} N_R [D^R_\mu,\R^R_\nu] 
+ \frac{1}{2}N_R F^R_{\mu\nu}
-\frac{p^2}{d}
[D^R_\mu,N_R][\R^R_\nu , N_R]
\right),
\label{eq.3.17}
\end{eqnarray}
where we have used $\R^R_\mu= \mlr^{-1}(D_\mu m)_{LR}$.

Some remarks are in order. In the right-hand side of (\ref{eq.3.17})
$\tr$ no longer includes Dirac space. Also, $d$ refers to the
dimension of momentum space in the sense of dimensional
regularization. This is a device related to Chan's formula which
applies to bosonic (Klein-Gordon) theories with arbitrary internal
degrees of freedom.  The Dirac gammas are all the time in two
dimensions (they are not dimensionally extended) and in particular
$\gamma_5=i\gamma_0\gamma_1$ anticommutes with all $\gamma_\mu$.

The first term inside the trace in (\ref{eq.3.17}) is not directly UV
convergent but it is so upon integration by parts. Similarly, the
identity
\begin{equation}
\int d^2x\epsilon_{\mu\nu}\tr\!\left(
\frac{1}{2}N_R F^R_{\mu\nu}-\frac{1}{2}N_L F^L_{\mu\nu}
\right)
=
\int d^2x\epsilon_{\mu\nu}\tr\!\Big(
-N_R(D_\mu m)_{RL}N_L(D_\nu m)_{LR}
\Big)
\end{equation}
allows to bring the second term to an UV convergent form (up to terms
which are symmetric under the exchange $L\leftrightarrow R$ and so
cancel in (\ref{eq:2.36a})). After those replacements $d$ can be set
to two:
\begin{eqnarray}
\Tr(\gamma_5\log K_R)_{d=2,\text{LO}} &=&
-i\int \frac{d^2x d^2p}{(2\pi)^2}\epsilon_{\mu\nu}\tr\!\Big(
 [D^R_\mu,N_R] \R^R_\nu 
\nonumber \\
&&
+N_R(D_\mu m)_{RL}N_L(D_\nu m)_{LR}
 +p^2 [D^R_\mu,N_R][\R^R_\nu , N_R]
\Big)
.
\end{eqnarray}

This can be worked out by using standard manipulations:
\begin{equation}
[D^R_\mu,N_R]= -N_R[D^R_\mu,\mrl\mlr]N_R,
\quad
[\R^R_\mu,N_R]= -N_R[\R^R_\mu,\mrl\mlr]N_R,
\end{equation}
plus integration by parts in momentum space to bring the expression to
a simpler form. Then the effective action can be written as
\begin{eqnarray}
W^-_{c,d=2,\text{LO}} &=&
-\frac{i}{2}\int \frac{d^2x d^2p}{(2\pi)^2}\epsilon_{\mu\nu}\tr\!\Big(
  N_R^2 \mrl (D_\mu m)_{LR}  N_R \mrl  (D_\nu m)_{LR}
\nonumber \\
&&
- N_L^2 \mlr (D_\mu m)_{RL}  N_L \mlr  (D_\nu m)_{RL}
\Big)
.
\label{eq:3.23}
\end{eqnarray}

Because the momentum appears in matricial expressions it is not
possible to carry out the momentum integral directly (except  for
particular abelian configurations). The obvious approach is then to
take matrix elements in an eigenbasis of the matrices involved. In the
present case one can define two (local) orthonormal basis of
eigenvectors in internal space, by the relations
\begin{equation}
\mlr|j,R\rangle=
m_j|j,L\rangle
,
\qquad
\mrl|j,L\rangle= m_j|j,R\rangle
,
\quad
m_j>0
.
\end{equation}
In fact the $|j,R\rangle$ are just eigenvectors of $\mrl\mlr$ with
eigenvalues $m_j^2$ while the $|j,L\rangle$ are eigenvectors of
$\mlr\mrl$ with the same eigenvalues $m_j^2$. One can now take matrix
elements to compute the trace in (\ref{eq:3.23}). In doing so
$\mrl\mlr$ and $\mlr\mrl$ are replaced by their eigenvalues and the
momentum integral becomes straightforward.  This gives\footnote{The
  limit $m_j^2\to m_i^2$ is finite and it correctly reproduces the
  contribution from diagonal matrix elements.}
\begin{eqnarray}
W^-_{c,d=2,\text{LO}} &=&
\frac{i}{8\pi}\int d^2x\,\epsilon_{\mu\nu} \sum_{i,j}
\frac{1}{m_i^2-m^2_j}\left[
\frac{1}{2}\left(\frac{1}{m_i^2}+\frac{1}{m_j^2}\right)
-\frac{\log(m^2_i/m^2_j)}{m^2_i-m^2_j}
\right]
\nonumber \\
&&
\times 
\Big(
\langle i,R|\mrl (D_\mu m)_{LR}|j,R\rangle
\langle j,R|\mrl (D_\nu m)_{LR} |i,R\rangle
\nonumber \\
&&
\qquad 
-
\langle i,L|\mlr (D_\mu m)_{RL}|j,L\rangle
\langle j,L|\mlr (D_\nu m)_{RL}|i,L\rangle
\Big)
.
\label{eq:3.24a}
\end{eqnarray}
This result is that already obtained in \cite{Salcedo:2000hx} where
the effective action is obtained from the current, a result later
verified in \cite{Hernandez:2007ng} using the world-line
approach,\footnote{In these two references the more efficient
  technique of labeled operators is used, instead of the eigenbasis
  method.} and should also be identical to that quoted in
\cite{Ball:1989xg} (p.146).

A similar calculation can be done for $W^+$. Here we find that the
result of \cite{Salcedo:2000hp} is also reproduced, after removal of a
spurious counter-term $1/(16\pi)\int d^2x\,\tr(\R^R_\mu
\R^R_\mu+\R^L_\mu \R^L_\mu) $.

The calculation of more complicated cases get quite involved using
this direct Chan's approach. This is due to the fact that $\gamma_\mu$
appears (twice) in the effective connection $B^{R,L}_\mu$. In
\cite{Salcedo:2008bs} we have set up a method of Chan type specific
for $K_{R,L}$ and have calculated $W^-_c$ to four derivatives in two
and four dimensions. It is found that the results quoted in
\cite{Salcedo:2000hx} for the LO in four dimensions and in
\cite{Hernandez:2007ng} for the NLO in two dimensions are correctly
reproduced.

\section{Conclusions}
\label{sec:conclusions}

In this work we have found a remarkable result, namely, there is a
local operator, $\Kop$, which correctly reproduces the non-trivial
chiral invariant factor of $\Det\Dirac$. Moreover, such operator is
 explicitly constructed. This allows to address a direct non
perturbative definition of the chiral determinant as well as it
evaluation with explicit chiral invariance throughout.

As nice feature is that, unlike other approaches, we have not
introduced chirally rotated fields to carry out our construction. We
work all the time with the original LR variables.\footnote{This is
  relevant since, as shown in \cite{Salcedo:2000hx}, the number of
  chiral invariant functionals depending {\em analytically} on the
  original variables that one can write down is much less than the
  total number of chiral invariant functionals (using, e.g., rotated
  variables).}  Another nice feature is that the calculation through
$\Tr\log\Kop$ allows to go directly to the terms one needs without
reconstruction from the current.

The derivation has been carried out for flat space-time, but at
present there is no indication that it cannot be extended to the
curved case, so we can conjecture that a similar operator $\Kop$
exists for curved space-times. This would allow to address the
computation of the effective action of fermions in the presence of
external gravitational fields, beyond the known WZW-like
contributions.  Likewise, the construction quite probably admits an
extension to manifolds with generic topology, and in particular those
corresponding to thermal compactification needed in the imaginary time
approach to finite temperature.

It is also interesting that bypassing the chiral obstruction suggests
a direct way to put the chiral invariant part of the chiral
determinant in the lattice. The geometrical (rather than dynamical)
anomalous $WZW$ term can then be added.

Finally, the construction presented here could be translatable to
other actions afflicted by anomalies. In particular, in the operator
formulation of non commutative field theory
\cite{Alvarez-Gaume:2000dx} it would seem that $\Kop$ would take the
same form as given here. This is because $x$ appears always in non
abelian fields for which we do not assume any particular commutation
properties. This observation is confirmed by the fact that the
heat-kernel expansion in non commutative field theory takes the same
form as in the ordinary one \cite{Vassilevich:2003yz}.

\acknowledgments
I thank C. Garc{\'\i}a-Recio for suggestions on the manuscript.
This work is supported in part by funds provided by the Spanish DGI
and FEDER funds with grant FIS2005-00810, Junta de Andaluc{\'\i}a
grants FQM225, FQM481 and P06-FQM-01735 and EU Integrated
Infrastructure Initiative Hadron Physics Project contract
RII3-CT-2004-506078.


\end{document}